\documentclass[prb,twocolumn,draft,amsmath,showpacs]{revtex4}
\usepackage{graphics}\input epsf

\begin{document}

\title{Orbiton-mediated  multi-phonon scattering in
La$_{1-x}$Sr$_x$MnO$_3$}

\author{K.-Y. Choi,$^{1,2}$
P. Lemmens,$^{3,4}$ G. G\"{u}ntherodt,$^2$ Yu. G. Pashkevich,$^5$
V. P. Gnezdilov,$^6$ P. Reutler,$^{7,8}$ L. Pinsard-Gaudart, $^8$
B. B\"{u}chner,$^7$ and A. Revcolevschi $^{8}$}

\affiliation{$^1$ Institute for Materials Research, Tohoku
University, Katahira 2-1-1, Sendai 980-8577, Japan}

\affiliation{$^2$ 2. Physikalisches Institut, RWTH Aachen, 52056
Aachen, Germany}

\affiliation{$^3$ Max Planck Institute for Solid State Research,
D-70569 Stuttgart, Germany}

\affiliation{$^4$ Institute for Physics of Condensed Matter, TU
Braunschweig, D-38106 Braunschweig, Germany}

\affiliation{$^5$ A. A. Galkin Donetsk Phystech NASU, 83114
Donetsk, Ukraine}

\affiliation{$^6$ B. I. Verkin Institute  for Low Temperature
Physics NASU, 61164 Kharkov, Ukraine}

\affiliation{$^7$ Institute for Solid State Research, IFW Dresden,
D-01171 Dresden, Germany}

\affiliation{$^8$ Laboratoire de Physico-Chimie, Universit\'{e}
Paris-Sud, 91405 Orsay, France}

\date{\today}

\begin{abstract}
We report on Raman scattering measurements of single crystalline
La$_{1-x}$Sr$_x$MnO$_3$ ($x$=0, 0.06, 0.09 and 0.125), focusing on
the high frequency regime. We observe  multi-phonon scattering
processes up to fourth-order which show distinct features: (i)
anomalies in peak energy and its relative intensity and (ii) a
pronounced temperature-, polarization-, and doping-dependence.
These features suggest a mixed orbiton-phonon nature of the
observed multi-phonon Raman spectra.
\end{abstract}
\pacs{71.45.Lr, 76.60.Gv, 61.72.Hh}


\maketitle


\section{Introduction}

Recently, there has been a debate about the underlying nature of
the experimentally reported orbitons  observed by Raman
spectroscopy in orbital ordered LaMnO$_3$.~\cite{SaitohN} Despite
pronounced temperature- and symmetry-dependent properties of the
modes, the assignment has been confronted with criticism and
alternative concepts. \cite{grueninger,saitoh-reply,Brink,Allen}
For instance, multi-phonon scattering has been theoretically
predicted to arise from the Franck-Condon (FC) process via
self-trapped orbitons.~\cite{Allen} Interestingly, in spite of
{\it mutually exclusive} selection rules an infrared absorption
study shows similar features  as Raman scattering measurements at
about 125 meV (1000~cm$\rm ^{-1}$), 145 meV (1160~cm$\rm ^{-1}$),
and 160 meV (1280~cm$\rm ^{-1}$).~\cite{grueninger} This was
attributed to multi-phonon scattering instead of orbital
excitations. On the other hand, orbiton-phonon mixed modes are
predicted to appear as satellite structures in the phonon spectrum
due to electron-phonon coupling.~\cite{Brink}

LaMnO$_3$ shows $d_{3x^2-r^2}$/$d_{3y^2-r^2}$ orbital ordering
below T$_{JT}=780$ K accompanied by a cooperative Jahn-Teller (JT)
distortion.~\cite{murakami} Two different contributions to this
antiferro-orbital ordering, which support each other, have been
suggested. The first mechanism relies on the cooperative JT
effect. In this case, orbiton excitations accompany lattice
distortions and {\it vice versa}. Their energy, determined by the
strength of electron-phonon coupling, is between 0.7 and
2~eV.~\cite{Allen} The second mechanism is based on superexchange
interaction. Within this mechanism orbitons are collective
excitations and their energy is expected at much lower energies of
160~meV because of strong on-site Coulomb
repulsion.~\cite{Okamoto02,Okamoto} Thus, the determination of the
respective contribution to orbital ordering and its energy scale
are closely related to the nature of orbital excitations.

Quite recently, resonant Raman scattering measurements
~\cite{Krueger} have shown a symmetry-dependent resonance of the
one- and two-phonon signal around the JT gap at 2 eV, suggesting a
phononic origin of the assigned orbitons due to the FC mechanism.
However, a detailed temperature- and symmetry-dependence is still
lacking which is significant in addressing a coupling of phonons
to orbital ordering. Thus, a careful examination on the higher
energy regime of the Raman spectrum above 1300~cm$\rm ^{-1}$ is
indispensable to uncover all aspects of orbital dynamics.
Furthermore, their doping dependence can give further clues
because the orbital ordering undergoes a transition while crossing
the phase boundary between the canted antiferromagnetic insulating
(CAF) and ferromagnetic insulating (FMI) state.~\cite{Geck}

In this paper we report on Raman scattering measurements of single
crystalline  La$_{1-x}$Sr$_x$MnO$_3$ ($x$=0, 0.06, 0.09 and
0.125), focusing on the higher energy regime. For $x=0$ we observe
multi-phonon scattering up to fourth order of the one-phonon modes
at frequencies between 490 and 640~cm$\rm ^{-1}$. The distinct
symmetry-, temperature-, and doping-dependencies of this
scattering contribution unveil the insufficiency of the FC
mechanism in describing the observed multi-phonon scattering. This
points to a orbiton-phonon mixed nature of the observed
multi-phonon scattering.


\begin{table}
\caption{\label{Table}Peak frequencies of multi-phonon scattering
of different orders. Multiples of first order scattering
frequencies are given in parentheses. The multi-phonon scattering
peaks are grouped according to their overtones of the 493-, 602-,
and 641-cm$^{-1}$ modes. Two different frequencies which belong to
the same group are attributed to different polarizations.}
\begin{ruledtabular}
\begin{tabular}{lllll}
1$^{st}$   & polarization &  2$^{nd}$  & 3$^{rd}$  & 4$^{th}$
\\\hline
493 & ($xx$)   & 985  (986)   & 1485 (1488)  &  -  (1972) \\
    & ($x'x'$) & 1017         & 1539         & 2096\\\hline
602 & ($xx$)   & 1137          & 1653         & 2297 \\
    & ($x'x'$) & 1173 (1204)  & 1781 (1806) &  2418 (2408)\\\hline
641 & ($xx,x'x'$)& 1281 (1282)  & 1911 (1923) &  2542 (2564)\\
\end{tabular}
\end{ruledtabular}
\end{table}

\section{Experimental details}

Single crystals of  La$_{1-x}$Sr$_x$MnO$_3$ ($x$=0, 0.06, 0.09,
and 0.125) were grown by using the floating zone method. These
samples are twinned. As a result, some crystallographic axes
cannot be discriminated.  Raman scattering measurements were
performed in a quasi-backscattering geometry with the excitation
line $\lambda= 514.5$ nm (2.34 eV) of an Ar$^{+}$ laser. The small
incident power of 8 mW avoids significant heating and irradiation
effects. Raman spectra were collected by a DILOR-XY triple
spectrometer and a nitrogen cooled CCD (charge-coupled device)
detector. The high temperature measurements from 300~K to 640~K
were carried out using a heating stage under vacuum.

\section{Experimental results}

Figure~1(a) displays polarized Raman spectra of LaMnO$_3$ at 5 K.
The incident and scattered light are parallel and perpendicular to
the quasicubic surface of the perovskitelike crystals. For such
scattering geometries, Raman spectra are expected in the $xx$,
$x'x'$, $xy$, and $x'y'$ polarizations where $x$, $y$, $x'$, and
$y'$ correspond to the respective quasicubic [100], [010], [110],
and [1$\bar{1}$0] directions. Since the studied samples are
twinned, the parallel configuration contains $xx$ and $x'x'$
polarizations while the crossed one consists of $xy$ and $x'y'$
polarizations.  Due to different selection rules the Raman spectra
of the parallel and crossed polarizations show a slightly
different behavior in the total number of phonon modes and their
relative intensity.

Below 650 ~cm$\rm ^{-1}$ we observe 17 phonon modes as first-order
scattering out of  24 symmetry-allowed modes for the $Pnma$
crystal structure. The number and sharpness of the observed modes
guarantees a high quality of the single crystal. The modes below
330 cm$^{-1}$ are due to vibrations of (La/Sr) cations and
rotations of the MnO$_6$ octahedra. The modes above 400 cm$^{-1}$
arise from bending and stretching vibrations of the
octahedra.~\cite{Reichardt} The atomic displacement of the
respective normal modes has been provided by M. N. Iliev {\it et
al.} [12]. For the detailed physics of the one-phonon modes we
refer to our previous work.~\cite{Choi} Hereafter, we will focus
on the high frequency regime covering the reported orbitons. Our
results reproduce well the three features at about 1000, 1160, and
1280~cm$\rm ^{-1}$ which were assigned to orbitons.~\cite{SaitohN}
Moreover, at much higher energies even additional maxima show up.
All maxima between 1485 and 1911~cm$\rm ^{-1}$ as well as between
2096 and 2542~cm$\rm ^{-1}$ appear close to periodically [see
Figs. 1(b) and (c)]. This suggests a common origin based on the
one-phonon modes between 493 and 641~cm$\rm ^{-1}$. Thus, the
claimed orbitons should not be considered as {\it pure} orbital
waves. In principle, a collective orbital wave might be also
observable besides multi-phonon scattering. However, we did not
find evidence for this collective state below 3000 cm$^{-1}$
($\approx 360$ meV). If present, this might be due to a negligible
scattering cross section caused by the strong JT distortion. This
implies that the orbital ordering in LaMnO$_3$ is predominantly
determined by the JT mechanism. A closer look at the higher-energy
maxima reveals a fine structure. This is due to a twinning of the
single crystal and the polarization-dependence of the multi-phonon
frequency. The latter feature is unusual in conventional
multi-phonon scattering and allows us to assign the symmetry of
the maxima by comparison to the Raman study of an untwinned single
crystal.~\cite{SaitohN}

\begin{figure}[t]
      \begin{center}
       \leavevmode
       \epsfxsize=8cm \epsfbox{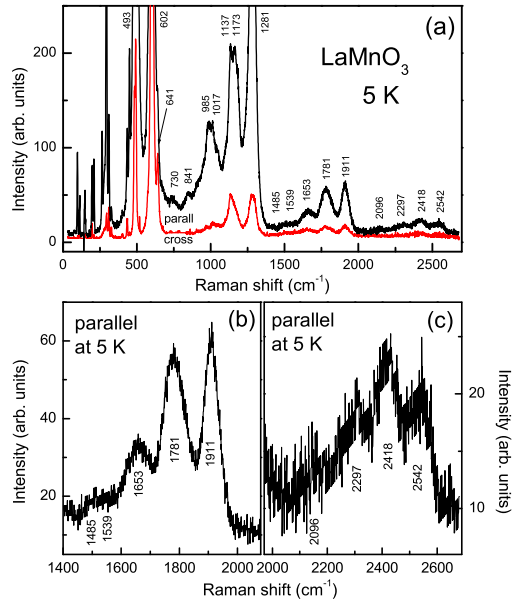}
        \caption{(online color) (a) Polarized Raman spectra of LaMnO$_3$ at 5 K in parallel (upper curve) and crossed (lower
curve) polarizations. Remarkably, multiphonon scatterings are
observed up to fourth order. (b) and (c) are a zoom on the modes
that correspond to scattering of third and fourth order,
respectively. } \label{fig.1}
\end{center}
\end{figure}

The measured frequencies of the high-energy excitations are listed
in Table I together with calculated integer multiples of
one-phonon modes. Here note that the assignment of two-phonon
peaks to specific one-phonon modes is not straightforward since
two-phonon scattering  arises predominantly from regions of the
Brillouin zone where the phonon density of states is largest.
Nonetheless, as a starting point we will assign them to simple
multiples of one-phonon peaks. This is because the FC mechanism is
asserted to be mainly responsible for the observed multi-phonon
scattering.~\cite{Krueger,Martin} In this local mechanism,
higher-order scattering usually shows up at integer multiples of
one-phonon peak energy, as the FC process results from a
displacement of the intermediate electronic state from the initial
one rather than a simultaneous emission of $n$-phonons. Thus, this
feature enables us to check the validity of the FC mechanism.
\begin{figure}[t]
      \begin{center}
       \leavevmode
       \epsfxsize=8cm \epsfbox{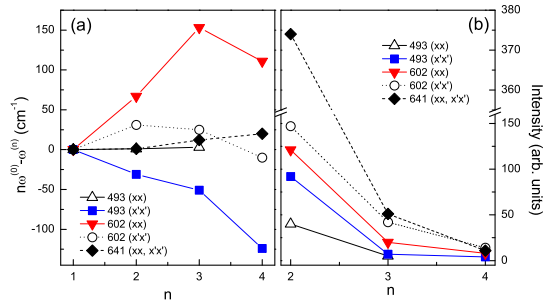}
        \caption{(online color) (a) The individual energy differences between $n$-phonon
scattering and integer multiples of one-phonon peaks as a function
of order. The energy difference is not equal to zero. In
particular, the overtones of the 493-cm$\rm ^{-1}$ ($x'x'$) and
602-cm$\rm ^{-1}$ ($xx$) modes show a linear behavior up to
third-order. A deviation from such a feature is seen at
fourth-order. (b) The integrated intensity of higher-order phonon
spectra. Especially, the 641-cm$\rm ^{-1}$ mode shows a strong
decrease of scattering intensity with increasing order.}
\label{fig.2}
\end{center}
\end{figure}

The 985- and 1485-cm$\rm ^{-1}$ peaks correspond to the overtones
of the 493-cm$\rm ^{-1}$ mode in $xx$ polarization. The 1017-,
1539-, and 2096-cm$\rm ^{-1}$ peaks might also be related to
higher-order scattering of the 493-cm$\rm ^{-1}$ mode in $x'x'$
polarization. Similarly, the 1137 (1173)-, 1653 (1781)-, and 2297
(2418)-cm$\rm ^{-1}$ modes are assigned to the overtones of the
602-cm$\rm ^{-1}$ mode in $xx$ ($x'x'$) polarization. The 1281-,
1911-, and 2542-cm$\rm ^{-1}$ modes correspond to the overtones of
the 641-cm$\rm ^{-1}$ mode. Several distinct features show up. The
second order scattering at 1281~cm$\rm ^{-1}$ is much more intense
than the first order one at 641~cm$\rm ^{-1}$. Furthermore, as
Fig. 2(a) displays, there are the energy differences between
multi-phonon scattering and integer multiples of one-phonon peaks.
For example, the overtones of the 602-cm$\rm ^{-1}$ mode in $xx$
polarization shift to higher energy up to third order and then
show a decrease at fourth order. In contrast, those of the
493-cm$\rm ^{-1}$ mode in $x'x'$ polarization shift to lower
energy with increasing order. The overtones of the 641-cm$\rm
^{-1}$ mode show no substantial anomaly in frequency. Instead, its
scattering intensity as a function of order does not parallel to
other modes contrary to the FC picture. As a consequence of a
strong decrease of scattering intensity at fourth order it becomes
weaker than that of the 602-cm$\rm ^{-1}$ mode in $x'x'$
polarization [see Fig. 1(c) and Fig. 2(b)]. This demonstrates that
the FC mechanism cannot capture the full aspect of the observed
multi-phonon scattering. Further note that even if one considers a
combination of first-order peaks, one cannot produce consistently
all higher-order peaks. \cite{AB}

\begin{figure}[t]
      \begin{center}
       \leavevmode
       \epsfxsize=6cm \epsfbox{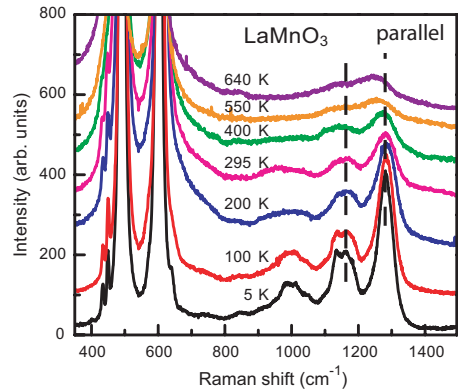}
        \caption{(online color) Polarized Raman spectra of LaMnO$_3$
         as a function of temperature. The spectra are
         shifted for clarity.} \label{fig.3}
\end{center}
\end{figure}

We turn now to the temperature dependence of the higher-order
Raman spectra. As Fig.~3 shows, upon heating the second order
maxima undergo a broadening and softening. The third and fourth
order signal (not shown here) can be detected up to 350 K and 150
K, respectively. Their behavior parallels that of the second order
signal within the measured temperature interval. The main features
are summarized in Fig.~4. First, with increasing temperature the
normalized intensity of the one-phonon mode  as well as of their
overtones falls off like an order parameter [see Fig.~4(a) for a
typical behavior at 493~cm$\rm ^{-1}$]. Second, the second order
mode  softens by 40~cm$\rm ^{-1}$ upon heating from 5 K to 640 K.
The representative example at 1281~cm$\rm ^{-1}$ is shown in
Fig.~4(a). This is contrasted by a small frequency shift of the
corresponding one-phonon mode by $\sim 6$~cm$\rm ^{-1}$.
Noticeably, its temperature dependence agrees well with that of
the phonon intensity. Third, the ratio of a second- to first-order
integrated phonon intensity, $I_{2}/I_{1}$, decreases gradually
with increasing temperature as Fig.~4(b) shows.

\begin{figure}[t]
      \begin{center}
       \leavevmode
       \epsfxsize=8cm \epsfbox{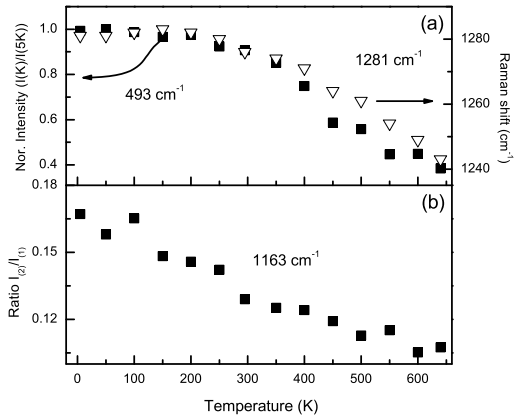}
        \caption{(a) Temperature dependence of the normalized intensity of the
493-cm$\rm ^{-1}$ mode as well as the peak energy of the
1281-cm$^{-1}$ mode. (b) The ratio of second- and first-order
scattering at 1163 cm$^{-1}$ as a function of temperature.}
\label{fig.4}
\end{center}
\end{figure}

Figure~5 displays a doping study of the second-order Raman spectra
of La$_{1-x}$Sr$_x$MnO$_3$ ($x$=0, 0.06, 0.09, and 0.125) at 5 K.
For the CAF samples ($x$=0, 0.06, and 0.09) similar features are
observed with increasing doping except a rapid decrease of the
intensity. In contrast, there exists only a broadened maximum
around 1000~cm$\rm ^{-1}$ in the FMI sample ($x$=0.125) while
maxima around 1160 and 1280 ~cm$\rm ^{-1}$ disappear. It should be
noted that the peak energies and relative intensities of the
corresponding one-phonon modes do not change
appreciably.~\cite{Choi} Furthermore, for $x=0.125$ the broadened
peak around 1000~cm$\rm ^{-1}$ should not be considered as the
reappearance of the peak which is already present in the CAF
samples. Rather, it should be ascribed to a damping of the maxima
of the CAF samples ranging from 1100 to 1350 ~cm$\rm ^{-1}$. This
is supported by the results of oxygen-doped manganites
LaMnO$_{3+\delta}$ ($0.071\leq \delta \leq 0.125$); the three
pronounced peaks in the CAF sample continuously evolve into a
broad, unstructured maximum in the FMI samples  which
systematically shifts to lower energy as $\delta$
increases.~\cite{Yurii}

\begin{figure}[t]
      \begin{center}
       \leavevmode
       \epsfxsize=6cm \epsfbox{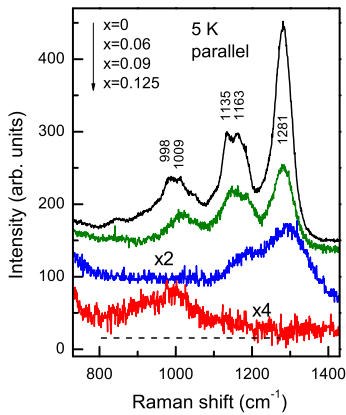}
        \caption{(online color) Second-order Raman spectra of
La$_{1-x}$Sr$_{x}$MnO$_3$  at 5 K as a function of doping ($x=$ 0,
0.06, 0.09, and 0.125)} \label{fig.5}
\end{center}
\end{figure}

In Fig.~6(a) we provide a polarization dependence of the $x=0.06$
sample owing to the discrimination between $xy$ plane and $z$
axis. Remarkably, in $zz$ polarization we observe solely the
1281-cm$\rm ^{-1}$ mode. This is quite unusual when taking into
account that the intensity of the 639-cm$\rm^{-1}$ mode is much
weaker than that of the 494-cm$\rm^{-1}$ mode. This symmetry
dependence reproduces partially the result reported in Ref.[1] and
can hardly be understood within a conventional multi-phonon
picture. Figures~6 (b) and (c) display the temperature dependence
for $x=0.09$ and 0.125. The $x=0.09$ shows a moderate decrease of
the second-order maxima upon heating from 5 K to room temperature.
In contrast, the FMI sample ($x=0.125$) exhibits a rather strong
suppression of the second-order signal through the metal-insulator
transition at T$_{C}\approx 185$ K  where the rearrangement of
orbital ordering takes place.~\cite{Geck} This suggests a
relationship between the observed multi-phonon scattering and the
change of an orbital ordering form through the CAF/FMI phase
boundary (see below).

\section{Discussions}

\begin{figure}[t]
      \begin{center}
       \leavevmode
       \epsfxsize=6cm \epsfbox{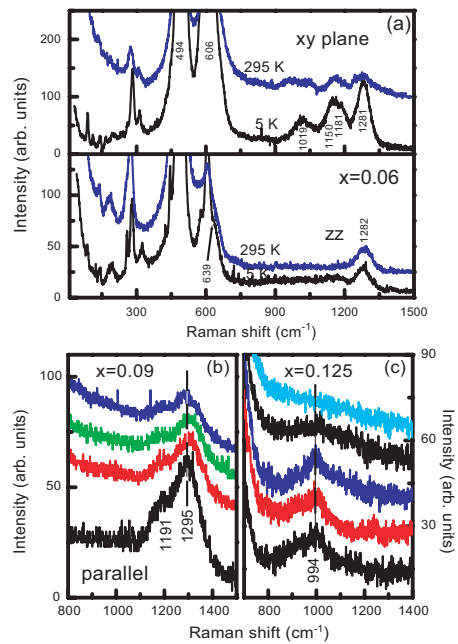}
        \caption{ (online color) (a) Raman spectra of La$_{0.94}$Sr$_{0.06}$MnO$_3$ with in plane
(upper panel) and $zz$ (lower panel) polarization at 5 K and room
temperature. (b) and (c) Temperature dependence of second-order
         Raman spectra of La$_{1-x}$Sr$_{x}$MnO$_3$; for $x=0.09$ at
         295, 200, 100, and 5 K and for $x=0.125$ at 295, 200, 185, 100, and
         5 K, respectively.} \label{fig.6}
\end{center}
\end{figure}

In the following we will examine the origin of the anomalous
multi-phonon scattering.

Usually, anharmonic electron-phonon interactions can lead to
multi-phonon scattering in transition metal oxides. The
simultaneous emission of $n$ phonons occurs due to lattice
anharmonicity. Since this mechanism is a simple extension of a
first-order process to higher order, the intensity of $n$-phonon
scattering decays as $g^{2n}$, where $g$ is the electron-phonon
coupling constant. In addition, the intensity ratio of a second-
to first-order scattering, that is $I_{2}/I_{1}$, would be
independent of temperature since the respective intensity has the
same resonant dependence on  incident photon energy.~\cite{Zhenya}
Obviously, these considerations cannot account for (i) the larger
intensity of the second-order peak at 1281~cm$\rm ^{-1}$ compared
to that of the first-order peak at 641~cm$\rm ^{-1}$
($I_{2}/I_{1}\sim 20$ at 5K), (ii) the anomalous evolution of peak
energies in higher-order scattering (see Table I and Fig.~2), and
(iii) the temperature-dependent ratio of $I_{2}/I_{1}$ [see Fig.
4(b)].

Another effective mechanism is based on resonance scattering. In
this process multi-phonons are created by a first-order
electron-lattice interaction together with a virtual electron
excitation.~\cite{Zhenya} When the incident photon energy is in
resonance with localized electronic excitations, multi-phonon
scattering becomes enhanced. Very recently, resonant Raman
scattering measurements of LaMnO$_3$ reveal  a sharp resonance
around 2 eV via the creation of {\it orbiton
excitons}.~\cite{Krueger} Since in our experiment the incident
light energy of 2.34~eV lies in the window of resonance (0.65 eV),
the observed multi-phonon scattering is also governed by resonant
scattering caused by orbitons. Significantly, orbital excitations
in the JT ordered state involve a local oxygen
displacement.~\cite{Allen} As a consequence, the FC mechanism is
expected to be active.

Actually, the validity of the FC mechanism  has been corroborated
by previous studies.~\cite{Krueger,Martin} Here we point out that
the FC mechanism is consistent with the simultaneous observation
of similar multi-phonon features by optical conductivity and Raman
scattering measurements despite completely different selection
rules.~\cite{grueninger} Originally, infrared inactive modes
become infrared active in the multi-phonon scattering owing to
nonlinear local oxygen displacements. In this case, the weak
intensity of the multi-phonon modes seen in the optical
conductivity can be ascribed to orbiton-assisted parity breaking.
However, as pointed out above, there exist several anomalies which
cannot be catched within the canonical FC picture.

First, the huge ratio of $I_{2}/I_{1}~\sim 20$ seen with respect
to the Mn-O stretching mode at 641-cm$\rm^{-1}$ is beyond  simple
theoretical calculations.~\cite{Allen,IRactive} Second, a
temperature-dependent ratio of $I_{2}/I_{1}$ is expected for the
resonant mechanism rather than for the FC one.~\cite{pantoja}
Third, the observed frequency and intensity of $n$-order
scattering (see Fig.~2) is incompatible  with the simple FC
scenario. We suggest that a strong coupling of orbitons to phonons
should also be taken into account.

Self-trapped orbital excitons accompany local ionic distortions,
that is, a strong nonlinearity corresponding to a superposition of
multi-phonons. As a result, orbitons are intrinsically coupled to
phonons under electron-phonon coupling. Theoretically, orbiton
induced satellites are predicted to develop at higher frequencies
of the phonon spectrum.~\cite{Brink} In this scenario, the
dominant energy scale of orbiton satellites is determined by the
strength of electron-phonon coupling. Noticeably, the anomalous
increase of the peak energy is observed at fourth-order scattering
between 2096 and 2542 cm$^{-1}$ [see Figs. 1 and 2]. If this
anomaly is considered to be a result of the influence of orbiton
satellites on multi-phonon scattering, one can obtain strong
electron-phonon coupling of $g\sim 1$.~\cite{Brink} However,
orbiton satellites do not show up as separate peaks. Rather, a
orbiton-phonon mixed character is reflected in diverse anomalies
of the observed multi-phonon modes.

This scenario can well explain the observed giant softening of the
two-phonon modes by 40 cm$^{-1}$ as a function of temperature [see
Fig.~4(a)]. Upon heating, the JT distortions become weaker as
demonstrated by the order-parameter-like fall-off of the intensity
of the Mn-O bond stretching mode in Fig.~4(a). The weakening of
electron-phonon coupling will lead to a shift of the orbiton
excitation energy to lower energies.~\cite{Brink} Further evidence
is provided by the temperature-dependent ratio of $I_{2}/I_{1}$.
With increasing temperature the local JT oxygen displacement
caused by the FC process will fade away. Consequently, the FC
contribution to multi-phonon scattering gradually decreases while
reducing the orbiton-phonon mixed character of multi-phonon
scattering. Then, a crossover of the FC mechanism to a
conventional one results in the reduced lifetime of the orbital
exciton seen in Fig.~4(b).

Also the drastic change of the two-phonon modes through the
CAF/FMI phase boundary supports our interpretation. The three
features of the CAF samples turn into  a broadened maximum around
1000~cm$\rm ^{-1}$ in the $x=0.125$ sample. The comparison between
the Sr- and oxygen-doped samples \cite{Yurii} unveils that the
broadened maximum around 1000~cm$\rm ^{-1}$ at $x=0.125$ results
from a softening of the maxima centered around 1300~cm$\rm ^{-1}$
at $x=0.09$. Such a giant shift of the two-phonon mode as a
function of doping is quite unusual and cannot be understood
without considering the orbital degrees of freedom. In the CAF
phase a $d_{3x^2-r^2}$/$d_{3y^2-r^2}$ type of an orbital ordering
is stabilized mainly by the cooperative JT distortions, leading to
the JT gap of 2 eV. When the incident light energy is within the
window of resonance (0.65 eV), Raman scattering process takes
place through an orbital flip from a $d_{3x^2-r^2}$/$d_{3y^2-r^2}$
orbital ordered state.~\cite{Krueger} In this resonant Raman
process, the energy and shape of multi-phonon scattering are
largely determined by a matrix element of orbiton excitons. Thus,
the three-peak feature of the observed two-phonon scattering and
their polarization dependence should be attributed to the specific
orbital ordering pattern of LaMnO$_3$. In the FMI phase a new type
of orbital state evolves from the LaMnO$_3$-type orbital state
upon cooling below T$_{C}\approx 185$ K while suppressing the JT
distortions.~\cite{Geck} Although its exact form is not known,
there exists evidence that orbital polarons are an essential part
of a new orbital ordered state.~\cite{Choi,Geck} Most probably,
thus, the orbital ordering of the FMI phase will be given as a
combination of an orbital polaron and a
$d_{3x^2-r^2}$/$d_{3y^2-r^2}$ orbital. Noticeably, the binding
energy of orbital polarons, $\Delta_{op}=0.6$ eV, is much smaller
than the JT gap of 2 eV.~\cite{Kilian99} As a result, Raman
scattering process will be mainly governed by an off-resonance
one. Then, this leads to a smearing of the features arising from
orbitons. Actually, the three-peak feature changes into the
unstructured maximum. In particular, this is related to a
fluctuation of the underlying orbitals in the FMI
phase.~\cite{Choi,Geck} The weakening of orbiton-phonon coupling
can account for a giant softening of the two-phonon frequency.
Therefore, we come to the conclusion that the multiphonon
scattering in LaMnO$_3$ relies strongly on the orbital ordered
pattern. Furthermore, the intrinsic orbiton-phonon couplings are
responsible for the observed anomalous behaviors in intensity and
frequency.

\section{Summary}

In summary, we have reported a detailed study of the higher-order
Raman scattering in the manganites La$_{1-x}$Sr$_x$MnO$_3$ as a
function of temperature and doping. For the first time, we are
able to probe and analyze phonon scattering up to forth-order
using Raman spectroscopy. Doping dependence of two-phonon
scattering together with the frequency shift of several higher
order modes indicates several anomalies which cannot be understood
within the canonical FC mechanism. The full understanding of the
multi-phonon scattering in the orbital-ordered LaMnO$_3$ system
can be achieved by considering the orbiton-phonon mixed character.
Our study suggests that in orbital ordered systems multi-phonon
scattering can serve as a valuable probe of the orbital dynamics.

\section*{Acknowledgements}

We thank J. Geck, R. Klingeler, C. Baumann, and M. Gr\"uninger for
useful discussions. This work was supported in part by the NATO
Collaborative Linkage Grant PST.CLG.977766 and INTAS 01-278 as
well as by DFG SPP1073.


\begin{thebibliography}{10}
\bibitem{IRactive}
Most probably, the 641-cm$\rm^{-1}$ mode is
infrared-active~\cite{grueninger}. Its observation by Raman
spectroscopy might be possible through defects/orbitons.
\end{thebibliography}
\end{document}